\documentclass[aps,prx,reprint]{revtex4-1}
\usepackage{blindtext}
\usepackage{graphicx}
\usepackage{xcolor}
\usepackage{dcolumn}
\usepackage{bm}
\usepackage[utf8]{inputenc}

\begin{document}
	
	\preprint{APS/123-QED}
	
	\title{Hidden electronic order in 1T-VSe$_{2}$}
	
	\author{T. Yilmaz}
	\affiliation{National Synchrotron Light Source II, Brookhaven National Lab, Upton, New York 11973, USA}
	
	\author{E. Vescovo}
	\affiliation{National Synchrotron Light Source II, Brookhaven National Lab, Upton, New York 11973, USA}

	\date{\today}
	
	\begin{abstract}
		Here, we study the surface electronic structure of 1T-VSe$_2$ by means of angle resolved photoemission spectroscopy and uncover a dispersion-less emission located in the vicinity of the Fermi level. Its crystal momentum dependency reveals that it occupies large portions of the Brillouin zone (BZ), where no bulk band is expected. Upon electron doping (deposition of Rb-atoms), the system evolves in a surprising way. Besides the expected down-shifting of the bands, a splitting of both bulk and the dispersion-less emission is observed. This peculiar behaviour strongly suggests the intrinsic nature of this emission. Its characterization may therefore be relevant to a deeper understanding of the physics of transition metal dichalcogenides.

	\end{abstract}

	\maketitle
	

	2D-layered, transition metal dichalcogenides (TMDCs) are ideally suited for advanced applications in electronics, spintronics, and optoelectronic \cite{mak2016photonics, manzeli20172d}. The large number of transition metals and chalcogens provides ample opportunities to synthesize a variety of electronic and crystal structures, hosting complex states of matter including superconductivity, Mott insulating states, topological insulators, ferromagnetism, and charge density wave states \cite{shi2015superconductivity, ritschel2015orbital, luxa2016origin, hsu2017topological}. Therefore, intense experimental and theoretical efforts have been applied to examine the band structure of these materials to reveal novel states of matter. All TMDCs in the 1T-crystal phase share a similar electronic structure, with only minor differences driven by the specific chemical potential \cite{chen2020strong, zeb2021tunable, woolley1977band}. Their Fermi surface topology is characterized by ellipsoidal electron pockets dominated by the $d$-orbitals of the transition metals and centered at the $\bar{M}$-points of the Brillion zone. The 4$p$-orbitals of the chalcogen atoms form instead highly dispersive bands which nearly touch the Fermi level at the $\bar{\Gamma}$-point.

	In spite of this simple band structure, TMDCs have occasionally been shown to support complex and exotic states \cite{chatterjee2015emergence}, whose careful experimental determination is therefore of primary importance. Here, we identify such hidden electronic order in the prototype dichalcogenide 1T-VSe$_2$. Our angle resolved photoemission spectroscopy (ARPES) experiments reveal that faint, dispersion-less electronic states, located in the vicinity of the Fermi level, cover the entire BZ. The faintness and dispersion-less character with a slowly decreasing density of state towards higher biding energy prevent these features to be clearly resolved in the 2D experimental maps. However, they can be distinguished in the Energy distribution curves (EDCs). Furthermore, surface deposition of the Rb atoms leads to splitting of this band and the bulk bands into multiple components. Similar band splitting is also seen in the V-shaped band dominated by V-3$d$ atomic orbitals. These observations indicate that this hidden electronic feature is correlated with the other bands and cannot be explained by extrinsic factors such as impurities or disorders. We also show that this spectral feature goes beyond the single-particle picture for this system. Our calculated band structure for 1T-VSe$_2$ is in agreement with previously published calculations and confirms the absence of this band close to the Fermi level in the computed band structure. Hence, this new finding could play an important role in the understanding and engineering of the unique electronic structure of TMDCs.
	
	Single crystal 1T-VSe$_2$ samples were obtained from 2dsemiconductors company (part number : BLK-VSe$_2$). The ARPES experiments were performed at the 21ID-I ESM beamline at the National Synchrotron Light Source II (NSLS-II), using a DA30 Scienta electron spectrometer with an energy resolution better than 15 meV and liner-horizontal polarized light. The pressure in the photoemission chamber was 3×10$^{-11}$ torr and samples were cooled by a closed cycle He cryostat and cleaved at 15 K just before collecting the photoemission data. Band structure calculations were performed with the Quantum Espresso (QE) package \cite{giannozzi2009quantum}, based on density functional theory without spin-orbit corrections.

	\begin{figure}[t]
		\centering
		\includegraphics[width=8.2cm,height=10.105cm]{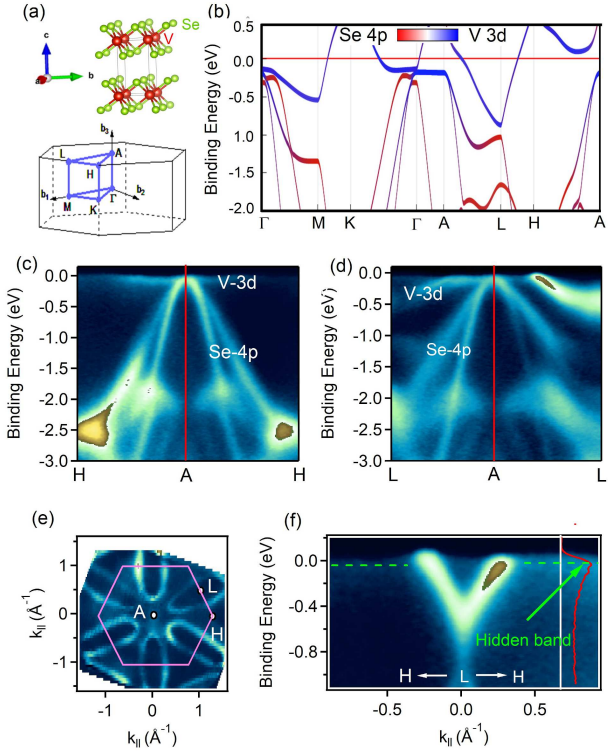}
		\caption{
			(a) Ball-and-stick representation of the bulk crystal structure of 1T-VSe$_2$ and corresponding hexagonal BZ. (b) Calculated band structure with atomic orbital characters. The straight horizontal line at 0 eV marks the Fermi level. (c) - (d) ARPES maps along the $A$ - $H$ and $A$ - $L$ directions. (e) Experimental Fermi surface of 1T-VSe$_2$. The superimposed hexagon identifies the 2D-BZ. (f) ARPES map along the $L$ - $H$ direction. The EDC away from the V-shaped band is also displayed (red curve) and its peak at the Fermi level is labeled as 'hidden band'. All spectra were recorded at 15 K with 169 eV photons, corresponding to k$_z$ $\simeq$ $A$-high symmetry point. 
		}
	\end{figure}
	
	The bulk crystal structure of 1T-VSe$_2$ can be visualized as a stack of 'monolayer'-building-blocks, weakly bonded by van der Waals (vdW) interactions (see Fig. 1 (a)). This system is therefore an ideal 2D material, each building-block comprising a V layer sandwiched between two Se layers. The hexagonal BZ for 1T-VSe$_2$ is also reproduced in Fig. 1(a) and the computed band structure is displayed in Fig. 1(b). The V and Se contributions to the valence band can be identified by the color coded orbital-character (red/blue for p/d-orbitals). Multiple hole bands, mostly of Se-4$p$ character (red), move towards the Fermi level on approaching the center of the BZ ($\Gamma$ and A high symmetry points), while the V-3d atomic orbitals (blue) forms relatively flat bands in the vicinity of the Fermi level.

	At first sight, the experimental band structure is highly consistent with the calculations. Strongly dispersive bands (Se-4$p$) and comparably flat bands close to E$_f$ (V-3$d$ orbitals) are indeed the main features seen in the ARPES maps (Fig. 1(c) and 1(d)). Furthermore, as expected, the experimental Fermi surface is  composed by ellipsoidal electron pockets centered at the $\bar{L}$-points (Fig. 1(e)). From the orbital projected band structure, it can be determined that these pockets are dominated by V-3$d$ atomic orbitals which form a V-shape band at the $L$-point (Fig. 1(f)). Thus, our ARPES study captures all the bands predicted by the computational method and these results are identical to earlier observations reported in the literature \cite{coelho2019charge, feng2018electronic}.
	
	Next, we focus on the regions of the electronic structure where no band is supposed to exit. One such region is located about the $H$-points, outside of the V-shape bulk band. Here, we observe a faint, dispersion-less feature marked with dashed green lines in Fig. 1(f). The interest in this emission, which is barely resolved in the ARPES maps, lies in its peculiar location, away from any predicted energy band (see the regions around the $K$ and $H$ - high symmetry points in the band structure of Fig. 1(b)). This feature is characterized by a spectral bump at the Fermi level with a slowly decreasing intensity towards higher binding energy as seen in the superimposed EDC in Fig. 1(f). The combination of low intensity, weak dispersion and high broadening makes the detection of this feature quite elusive and can explain why it has not been reported in earlier ARPES studies \cite{coelho2019charge, feng2018electronic, chen2018unique}. Furthermore, the spectral line-shape of this dispersion-less band does not exhibit the characteristic behavior of background intensity which would be expected to decrease towards the Fermi level. Therefore, this band, in the following referred to as a hidden band, is intrinsic to the material rather than induced by structural imperfections or impurities.
	
	\begin{figure}
		\centering
		\includegraphics[width=8.2cm,height=7.747cm]{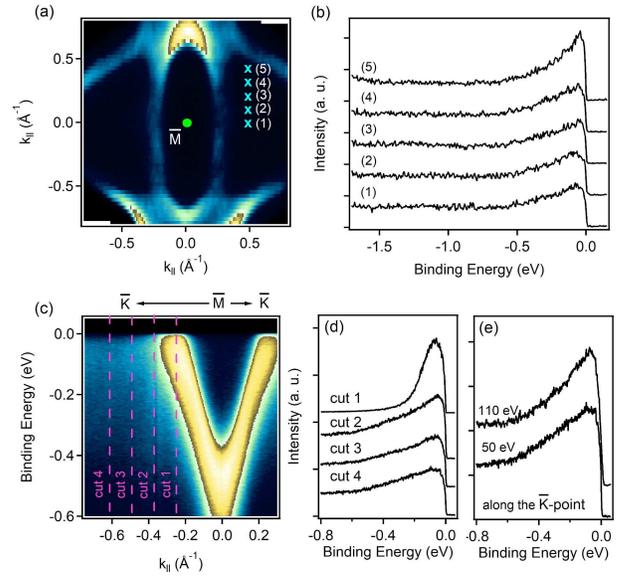}
		
		\caption{(a) Portion of the Fermi surface centered at the $\bar{M}$-point. (b) EDCs taken at the momentum points marked in (a). (c) ARPES map centered at the $\bar{M}$-point. (d) EDCs along the dashed pink lines in (c). (e) Photon energy dependent EDCs for identical momentum cuts. The ARPES maps in (a) and (c) were recorded at 15 K with 50 eV photons.}
		
	\end{figure}
	
	To further validate our observation, we investigate the energy dispersion along various momentum directions of the 3D-BZ. Several EDCs taken at the marked momentum points in Fig. 2(a) are presented in Fig. 2(b). These points are aligned parallel to the $\bar{M}$ - $\bar{\Gamma}$ direction and are chosen within the region where no band is expected in the vicinity of the Fermi level. In contrast, all EDCs exhibit a peak in the vicinity of the Fermi level with similar line-shape and intensity. Nominally, V-3$d$ and Se-4$p$ atomic orbitals are strongly dispersive (Fig. 1(b) -1(c)) implying that this feature is not due to secondary electrons from scattering which would mimic the primary electronic dispersion. Similarly, the EDCs taken at various k-points on the $\bar{M}$ - $\bar{K}$ direction, obtained from the ARPES map given in Fig.2(c), display the same spectral line-shape and intensity (Fig. 2(d)). For comparison, in Fig. 2 (d), we also show the EDC taken along the momentum point at the tip of the V-shape bulk band. At the Fermi level, its intensity is only twice the intensity of the hidden band. However, the spectral shape of the hidden band spreads more towards the higher binding energy which could indicate the presence of multiple peaks. In summary, having analyzed two perpendicular directions in the momentum space, we conclude that the hidden band probably covers the entire 2D-BZ.
	
	In principle, a weak, dispertion-less spectral feature located in the vicinity of the Fermi and spread over a large portion of the BZ could be attributed to impurity induced states \cite{kang2003resonant}. To check this possibility, we investigate the EDCs taken with 50 eV, close to the resonant V 3$p$-3$d$ transition and with 110 eV photons, far away from resonanace (Fig. 2(e)). Clearly, the spectral feature of the hidden band does not exhibit any prominent dependence on the photon energy, strongly indicating that it does not originate from impurity bands.
	
	\begin{figure}
		\centering
		\includegraphics[width=8.2cm,height=5.999cm]{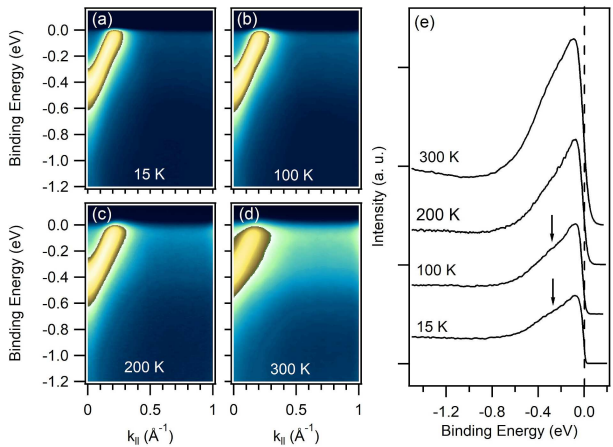}
		\caption{(a) - (d) Temperature dependency: ARPES maps from 1T-VSe$_2$ along the $\bar{M}$ - $\bar{K}$ direction. (b) MDCs at $k_\Vert$ = 0.6 $\AA^{-1}$. Spectra were collected with 92.5 eV photons.}
	\end{figure}
	
	We also consider the effect of temperature on the hidden band. Fig. 3(a) - 3(d) display the ARPES maps of 1T-VSe$_2$ taken at 15 K, 100 K, 200 K, and 300 K along the $\bar{M}$  - $\bar{K}$ direction. The EDCs obtained from these spectra at $k_\Vert$ = 0.6 $\AA^{-1}$ are collected in Fig. 3(e). In these high statistic data, the hidden band clearly reveals its composite nature, with an evident shoulder at higher binding energies, particularly at 15 and 100 K (see arrows in Fig. 3(e)) and less so at higher temperatures, due to thermal broadening. Another interesting observation is that the density of states in the hidden band increases with temperature, making its observation quite distinct even in the bare ARPES maps of Fig. 3(e) and 3(d). Such a spectral evolution with temperature might indicate a strong electron-phonon coupling for the hidden band.
	
	\begin{figure}
		\centering
		\includegraphics[width=8.2cm,height=9.368cm]{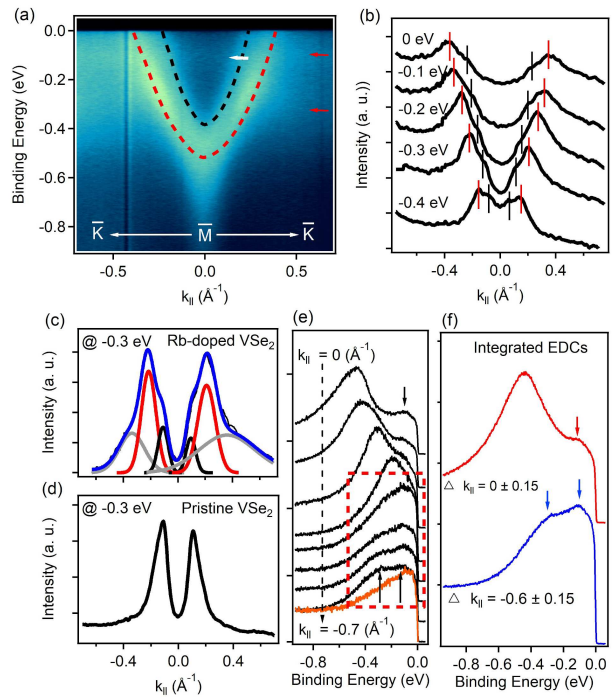}
		\caption{Rb-deposition: (a) ARPES map from Rb-deposited 1T-VSe$_2$ along the $\bar{M}$ - $\bar{K}$ direction. Black and red parabolas are guides to the eye. (b) Momentum distribution curves (MDC) at the indicated binding energies. Black and red vertical lines trace the two bands highlighted in (a). (c) MDC at -0.3 eV binding energy. Black, red, and gray doublets are Voigt fitting profiles. The overall fitting is the blue curve. (d) MDC at -0.3 eV from pristine 1T-VSe$_2$. (e) EDCs as a function of momentum. The red spectrum is from the pristine sample for comparison. The red dashed square focuses on the hidden band splitting where upward arrows marks the two componenets. Downward arrow marks the band indicated with white arrow in (a). (f) Integrated EDCs as specified in the fig. Blue and red arrows mark the band locations. Spectra are recorded with 50 eV photons at 15 K.}
	\end{figure}

	The broad line-shape of the hidden band suggests the presence of overlapping peaks. By breaking the inversion symmetry at the surface, it may be possible to further lift the spin degeneracy or inducing negative electron compressibility in these states \cite{riley2015negative, kang2017universal}, thereby making it easier to resolve any underlying substructure. To this end, in the next step, we study the effect of deposition of a submonolayer of Rb on the surface electronic structure of 1T-VSe$_2$. Our results are shown in Fig. 4. Surprisingly, instead of a simple band shift due to electron doping, the effect of Rb deposition is a striking modification of the electronic structure. The ARPES map taken along the $\bar{K}$ - $\bar{M}$ - $\bar{K}$ is shown in Fig. 4(a). The first thing to notice is that the original V-shaped band is now split into multiple components marked with red and black dashed parabolas. The band splitting can be easily appreciated in the MDC sections taken at different binding energies. Two distinct peaks are marked with vertical solid lines (red and black, Fig. 4(b)). To better visualize the band splitting, the MDC taken at -0.3  eV binding energy is fitted with Voigt profiles, resolving three distinct features (Fig. 4(c)). The inner two peaks are possibly due to the lifted spin degeneracy while the origin of the outer band remains elusive in our work. On the other hand, the identical MDC taken from the pristine sample exhibits only a single doublet without evident intra-band structure.

	Similarly, Rb deposition induces a band splitting in the hidden band. Even though this is not very clear in the ARPES map given in Fig. 4(a) (see red horizontal arrows), two weak peaks with an energy separation of 0.17 eV marked with upward arrows in Fig. 4(e) can be identified in the EDCs. Compared to the spectra taken from the pristine sample (red spectra in Fig. 4(e)), an increased intensity in the higher binding energy side of the EDC is observed for the Rb-deposited sample. This further supports the picture of multiple electronic states in the hidden band. Surprisingly, another weak band is also observable inside the V shape band located at 0.1 eV binding energy. This can be seen in the EDC taken at the $\bar{M}$-point. All these new spectral features are more evident in the EDCs integrated over a relatively wide momentum window (Fig. 4(f)). In conclusion, Rb deposition induces simultaneous splitting in the hidden band and the V-shape bulk band, further suggesting the intrinsic nature of the hidden band.
	
	To the best of our knowledge, such faint and dispersion-less spectral feature, located in the vicinity of the Fermi level, has been predicted only in materials supporting Holstein polarons, characterized by electron-phonon coupling limited within few atoms (short-range) or several atoms (long-rage) \cite{li2021general, fetherolf2020unification, berciu2008light}. From the band structure point of view, these excitation should manifest as splitting of a continuum parabolic band into sub-bands separated by energy gaps. In close analogy to the hidden band reported here, the edge of these bands display dispersion-less tails spreading over large portions of the BZ. Furthermore, in the Holstein polaron model, a band flattening at the bottom of the sub-bands was also predicted and later confirmed in 2H-MoS$_2$ by impurity deposition on the sample surface \cite{kang2018holstein}. Similarly, we observe the formation of an additional flat feature inside the V-shape band after the surface deposition of the Rb-impurities (Fig. 4(a)).
	
	Within this scenario, 1T-VSe$_2$ can be expected to be proximal to superconducting. Remarkably, this has been found to be the case. A recent work reports on the emergence of superconductivity in 1T-VSe$_2$ under intense quasi-hydrostatic pressure \cite{sahoo2020pressure}. The critical temperature for superconductivity is as high as $\sim$4 K at $\sim$15 GPa, which is significantly higher than in any other 1T-TMDC. In this study, the origin of the superconductivity is attributed to fine-tuning of the chemical potential.
	
	Finally, one of the known characteristics of polarons is the weakness of the quasi-particle peak at the Fermi level \cite{kang2018holstein, mannella2007polaron}. Therefore, a direct experimental proof of such electron-phonon coupling may be extremely difficult to obtain. Possibly ultra-high resolution ARPES experiments could be devised to test this scenario. In this connection, we hope that our work can serve as a significant contribution to inspire further investigations aiming at a deeper understanding of the unique electronic anomalies observed in TMDCs and ultimately at better controlling their physical properties.

	This research also used resources ESM (21ID-I) beamline of the National Synchrotron Light Source II, a U.S. Department of Energy (DOE) Office of Science User Facility operated for the DOE Office of Science by Brookhaven National Laboratory under Contract No. DE-SC0012704. We have no conflict of
	interest, financial or other to declare.

\bibliographystyle{apsrev4-1} 

\begin{thebibliography}{23}%
\makeatletter
\providecommand \@ifxundefined [1]{%
 \@ifx{#1\undefined}
}%
\providecommand \@ifnum [1]{%
 \ifnum #1\expandafter \@firstoftwo
 \else \expandafter \@secondoftwo
 \fi
}%
\providecommand \@ifx [1]{%
 \ifx #1\expandafter \@firstoftwo
 \else \expandafter \@secondoftwo
 \fi
}%
\providecommand \natexlab [1]{#1}%
\providecommand \enquote  [1]{``#1''}%
\providecommand \bibnamefont  [1]{#1}%
\providecommand \bibfnamefont [1]{#1}%
\providecommand \citenamefont [1]{#1}%
\providecommand \href@noop [0]{\@secondoftwo}%
\providecommand \href [0]{\begingroup \@sanitize@url \@href}%
\providecommand \@href[1]{\@@startlink{#1}\@@href}%
\providecommand \@@href[1]{\endgroup#1\@@endlink}%
\providecommand \@sanitize@url [0]{\catcode `\\12\catcode `\$12\catcode
  `\&12\catcode `\#12\catcode `\^12\catcode `\_12\catcode `\%12\relax}%
\providecommand \@@startlink[1]{}%
\providecommand \@@endlink[0]{}%
\providecommand \url  [0]{\begingroup\@sanitize@url \@url }%
\providecommand \@url [1]{\endgroup\@href {#1}{\urlprefix }}%
\providecommand \urlprefix  [0]{URL }%
\providecommand \Eprint [0]{\href }%
\providecommand \doibase [0]{http://dx.doi.org/}%
\providecommand \selectlanguage [0]{\@gobble}%
\providecommand \bibinfo  [0]{\@secondoftwo}%
\providecommand \bibfield  [0]{\@secondoftwo}%
\providecommand \translation [1]{[#1]}%
\providecommand \BibitemOpen [0]{}%
\providecommand \bibitemStop [0]{}%
\providecommand \bibitemNoStop [0]{.\EOS\space}%
\providecommand \EOS [0]{\spacefactor3000\relax}%
\providecommand \BibitemShut  [1]{\csname bibitem#1\endcsname}%
\let\auto@bib@innerbib\@empty
\bibitem [{\citenamefont {Mak}\ and\ \citenamefont
  {Shan}(2016)}]{mak2016photonics}%
  \BibitemOpen
  \bibfield  {author} {\bibinfo {author} {\bibfnamefont {K.~F.}\ \bibnamefont
  {Mak}}\ and\ \bibinfo {author} {\bibfnamefont {J.}~\bibnamefont {Shan}},\
  }\href@noop {} {\bibfield  {journal} {\bibinfo  {journal} {Nature Photonics}\
  }\textbf {\bibinfo {volume} {10}},\ \bibinfo {pages} {216} (\bibinfo {year}
  {2016})}\BibitemShut {NoStop}%
\bibitem [{\citenamefont {Manzeli}\ \emph {et~al.}(2017)\citenamefont
  {Manzeli}, \citenamefont {Ovchinnikov}, \citenamefont {Pasquier},
  \citenamefont {Yazyev},\ and\ \citenamefont {Kis}}]{manzeli20172d}%
  \BibitemOpen
  \bibfield  {author} {\bibinfo {author} {\bibfnamefont {S.}~\bibnamefont
  {Manzeli}}, \bibinfo {author} {\bibfnamefont {D.}~\bibnamefont
  {Ovchinnikov}}, \bibinfo {author} {\bibfnamefont {D.}~\bibnamefont
  {Pasquier}}, \bibinfo {author} {\bibfnamefont {O.~V.}\ \bibnamefont
  {Yazyev}}, \ and\ \bibinfo {author} {\bibfnamefont {A.}~\bibnamefont {Kis}},\
  }\href@noop {} {\bibfield  {journal} {\bibinfo  {journal} {Nature Reviews
  Materials}\ }\textbf {\bibinfo {volume} {2}},\ \bibinfo {pages} {1} (\bibinfo
  {year} {2017})}\BibitemShut {NoStop}%
\bibitem [{\citenamefont {Shi}\ \emph {et~al.}(2015)\citenamefont {Shi},
  \citenamefont {Ye}, \citenamefont {Zhang}, \citenamefont {Suzuki},
  \citenamefont {Yoshida}, \citenamefont {Miyazaki}, \citenamefont {Inoue},
  \citenamefont {Saito},\ and\ \citenamefont
  {Iwasa}}]{shi2015superconductivity}%
  \BibitemOpen
  \bibfield  {author} {\bibinfo {author} {\bibfnamefont {W.}~\bibnamefont
  {Shi}}, \bibinfo {author} {\bibfnamefont {J.}~\bibnamefont {Ye}}, \bibinfo
  {author} {\bibfnamefont {Y.}~\bibnamefont {Zhang}}, \bibinfo {author}
  {\bibfnamefont {R.}~\bibnamefont {Suzuki}}, \bibinfo {author} {\bibfnamefont
  {M.}~\bibnamefont {Yoshida}}, \bibinfo {author} {\bibfnamefont
  {J.}~\bibnamefont {Miyazaki}}, \bibinfo {author} {\bibfnamefont
  {N.}~\bibnamefont {Inoue}}, \bibinfo {author} {\bibfnamefont
  {Y.}~\bibnamefont {Saito}}, \ and\ \bibinfo {author} {\bibfnamefont
  {Y.}~\bibnamefont {Iwasa}},\ }\href@noop {} {\bibfield  {journal} {\bibinfo
  {journal} {Scientific reports}\ }\textbf {\bibinfo {volume} {5}},\ \bibinfo
  {pages} {1} (\bibinfo {year} {2015})}\BibitemShut {NoStop}%
\bibitem [{\citenamefont {Ritschel}\ \emph {et~al.}(2015)\citenamefont
  {Ritschel}, \citenamefont {Trinckauf}, \citenamefont {Koepernik},
  \citenamefont {B{\"u}chner}, \citenamefont {Zimmermann}, \citenamefont
  {Berger}, \citenamefont {Joe}, \citenamefont {Abbamonte},\ and\ \citenamefont
  {Geck}}]{ritschel2015orbital}%
  \BibitemOpen
  \bibfield  {author} {\bibinfo {author} {\bibfnamefont {T.}~\bibnamefont
  {Ritschel}}, \bibinfo {author} {\bibfnamefont {J.}~\bibnamefont {Trinckauf}},
  \bibinfo {author} {\bibfnamefont {K.}~\bibnamefont {Koepernik}}, \bibinfo
  {author} {\bibfnamefont {B.}~\bibnamefont {B{\"u}chner}}, \bibinfo {author}
  {\bibfnamefont {M.~v.}\ \bibnamefont {Zimmermann}}, \bibinfo {author}
  {\bibfnamefont {H.}~\bibnamefont {Berger}}, \bibinfo {author} {\bibfnamefont
  {Y.~I.}\ \bibnamefont {Joe}}, \bibinfo {author} {\bibfnamefont
  {P.}~\bibnamefont {Abbamonte}}, \ and\ \bibinfo {author} {\bibfnamefont
  {J.}~\bibnamefont {Geck}},\ }\href@noop {} {\bibfield  {journal} {\bibinfo
  {journal} {Nature physics}\ }\textbf {\bibinfo {volume} {11}},\ \bibinfo
  {pages} {328} (\bibinfo {year} {2015})}\BibitemShut {NoStop}%
\bibitem [{\citenamefont {Luxa}\ \emph {et~al.}(2016)\citenamefont {Luxa},
  \citenamefont {Jankovsk{\`y}}, \citenamefont {Sedmidubsk{\`y}}, \citenamefont
  {Medl{\'\i}n}, \citenamefont {Mary{\v{s}}ko}, \citenamefont {Pumera},\ and\
  \citenamefont {Sofer}}]{luxa2016origin}%
  \BibitemOpen
  \bibfield  {author} {\bibinfo {author} {\bibfnamefont {J.}~\bibnamefont
  {Luxa}}, \bibinfo {author} {\bibfnamefont {O.}~\bibnamefont {Jankovsk{\`y}}},
  \bibinfo {author} {\bibfnamefont {D.}~\bibnamefont {Sedmidubsk{\`y}}},
  \bibinfo {author} {\bibfnamefont {R.}~\bibnamefont {Medl{\'\i}n}}, \bibinfo
  {author} {\bibfnamefont {M.}~\bibnamefont {Mary{\v{s}}ko}}, \bibinfo {author}
  {\bibfnamefont {M.}~\bibnamefont {Pumera}}, \ and\ \bibinfo {author}
  {\bibfnamefont {Z.}~\bibnamefont {Sofer}},\ }\href@noop {} {\bibfield
  {journal} {\bibinfo  {journal} {Nanoscale}\ }\textbf {\bibinfo {volume}
  {8}},\ \bibinfo {pages} {1960} (\bibinfo {year} {2016})}\BibitemShut
  {NoStop}%
\bibitem [{\citenamefont {Hsu}\ \emph {et~al.}(2017)\citenamefont {Hsu},
  \citenamefont {Vaezi}, \citenamefont {Fischer},\ and\ \citenamefont
  {Kim}}]{hsu2017topological}%
  \BibitemOpen
  \bibfield  {author} {\bibinfo {author} {\bibfnamefont {Y.-T.}\ \bibnamefont
  {Hsu}}, \bibinfo {author} {\bibfnamefont {A.}~\bibnamefont {Vaezi}}, \bibinfo
  {author} {\bibfnamefont {M.~H.}\ \bibnamefont {Fischer}}, \ and\ \bibinfo
  {author} {\bibfnamefont {E.-A.}\ \bibnamefont {Kim}},\ }\href@noop {}
  {\bibfield  {journal} {\bibinfo  {journal} {Nature communications}\ }\textbf
  {\bibinfo {volume} {8}},\ \bibinfo {pages} {1} (\bibinfo {year}
  {2017})}\BibitemShut {NoStop}%
\bibitem [{\citenamefont {Chen}\ \emph {et~al.}(2020)\citenamefont {Chen},
  \citenamefont {Ruan}, \citenamefont {Wu}, \citenamefont {Tang}, \citenamefont
  {Ryu}, \citenamefont {Tsai}, \citenamefont {Lee}, \citenamefont {Kahn},
  \citenamefont {Liou}, \citenamefont {Jia} \emph {et~al.}}]{chen2020strong}%
  \BibitemOpen
  \bibfield  {author} {\bibinfo {author} {\bibfnamefont {Y.}~\bibnamefont
  {Chen}}, \bibinfo {author} {\bibfnamefont {W.}~\bibnamefont {Ruan}}, \bibinfo
  {author} {\bibfnamefont {M.}~\bibnamefont {Wu}}, \bibinfo {author}
  {\bibfnamefont {S.}~\bibnamefont {Tang}}, \bibinfo {author} {\bibfnamefont
  {H.}~\bibnamefont {Ryu}}, \bibinfo {author} {\bibfnamefont {H.-Z.}\
  \bibnamefont {Tsai}}, \bibinfo {author} {\bibfnamefont {R.~L.}\ \bibnamefont
  {Lee}}, \bibinfo {author} {\bibfnamefont {S.}~\bibnamefont {Kahn}}, \bibinfo
  {author} {\bibfnamefont {F.}~\bibnamefont {Liou}}, \bibinfo {author}
  {\bibfnamefont {C.}~\bibnamefont {Jia}},  \emph {et~al.},\ }\href@noop {}
  {\bibfield  {journal} {\bibinfo  {journal} {Nature Physics}\ }\textbf
  {\bibinfo {volume} {16}},\ \bibinfo {pages} {218} (\bibinfo {year}
  {2020})}\BibitemShut {NoStop}%
\bibitem [{\citenamefont {Zeb}\ \emph {et~al.}(2021)\citenamefont {Zeb},
  \citenamefont {Zhao}, \citenamefont {Ullah}, \citenamefont {Menezes},\ and\
  \citenamefont {Zhang}}]{zeb2021tunable}%
  \BibitemOpen
  \bibfield  {author} {\bibinfo {author} {\bibfnamefont {J.}~\bibnamefont
  {Zeb}}, \bibinfo {author} {\bibfnamefont {X.}~\bibnamefont {Zhao}}, \bibinfo
  {author} {\bibfnamefont {S.}~\bibnamefont {Ullah}}, \bibinfo {author}
  {\bibfnamefont {M.~G.}\ \bibnamefont {Menezes}}, \ and\ \bibinfo {author}
  {\bibfnamefont {W.}~\bibnamefont {Zhang}},\ }\href@noop {} {\bibfield
  {journal} {\bibinfo  {journal} {Journal of Materials Science}\ }\textbf
  {\bibinfo {volume} {56}},\ \bibinfo {pages} {6891} (\bibinfo {year}
  {2021})}\BibitemShut {NoStop}%
\bibitem [{\citenamefont {Woolley}\ and\ \citenamefont
  {Wexler}(1977)}]{woolley1977band}%
  \BibitemOpen
  \bibfield  {author} {\bibinfo {author} {\bibfnamefont {A.~M.}\ \bibnamefont
  {Woolley}}\ and\ \bibinfo {author} {\bibfnamefont {G.}~\bibnamefont
  {Wexler}},\ }\href@noop {} {\bibfield  {journal} {\bibinfo  {journal}
  {Journal of Physics C: Solid State Physics}\ }\textbf {\bibinfo {volume}
  {10}},\ \bibinfo {pages} {2601} (\bibinfo {year} {1977})}\BibitemShut
  {NoStop}%
\bibitem [{\citenamefont {Chatterjee}\ \emph {et~al.}(2015)\citenamefont
  {Chatterjee}, \citenamefont {Zhao}, \citenamefont {Iavarone}, \citenamefont
  {Di~Capua}, \citenamefont {Castellan}, \citenamefont {Karapetrov},
  \citenamefont {Malliakas}, \citenamefont {Kanatzidis}, \citenamefont {Claus},
  \citenamefont {Ruff} \emph {et~al.}}]{chatterjee2015emergence}%
  \BibitemOpen
  \bibfield  {author} {\bibinfo {author} {\bibfnamefont {U.}~\bibnamefont
  {Chatterjee}}, \bibinfo {author} {\bibfnamefont {J.}~\bibnamefont {Zhao}},
  \bibinfo {author} {\bibfnamefont {M.}~\bibnamefont {Iavarone}}, \bibinfo
  {author} {\bibfnamefont {R.}~\bibnamefont {Di~Capua}}, \bibinfo {author}
  {\bibfnamefont {J.~P.}\ \bibnamefont {Castellan}}, \bibinfo {author}
  {\bibfnamefont {G.}~\bibnamefont {Karapetrov}}, \bibinfo {author}
  {\bibfnamefont {C.~D.}\ \bibnamefont {Malliakas}}, \bibinfo {author}
  {\bibfnamefont {M.~G.}\ \bibnamefont {Kanatzidis}}, \bibinfo {author}
  {\bibfnamefont {H.}~\bibnamefont {Claus}}, \bibinfo {author} {\bibfnamefont
  {J.~P.~C.}\ \bibnamefont {Ruff}},  \emph {et~al.},\ }\href@noop {} {\bibfield
   {journal} {\bibinfo  {journal} {Nature communications}\ }\textbf {\bibinfo
  {volume} {6}},\ \bibinfo {pages} {1} (\bibinfo {year} {2015})}\BibitemShut
  {NoStop}%
\bibitem [{\citenamefont {Giannozzi}\ \emph {et~al.}(2009)\citenamefont
  {Giannozzi}, \citenamefont {Baroni}, \citenamefont {Bonini}, \citenamefont
  {Calandra}, \citenamefont {Car}, \citenamefont {Cavazzoni}, \citenamefont
  {Ceresoli}, \citenamefont {Chiarotti}, \citenamefont {Cococcioni},
  \citenamefont {Dabo} \emph {et~al.}}]{giannozzi2009quantum}%
  \BibitemOpen
  \bibfield  {author} {\bibinfo {author} {\bibfnamefont {P.}~\bibnamefont
  {Giannozzi}}, \bibinfo {author} {\bibfnamefont {S.}~\bibnamefont {Baroni}},
  \bibinfo {author} {\bibfnamefont {N.}~\bibnamefont {Bonini}}, \bibinfo
  {author} {\bibfnamefont {M.}~\bibnamefont {Calandra}}, \bibinfo {author}
  {\bibfnamefont {R.}~\bibnamefont {Car}}, \bibinfo {author} {\bibfnamefont
  {C.}~\bibnamefont {Cavazzoni}}, \bibinfo {author} {\bibfnamefont
  {D.}~\bibnamefont {Ceresoli}}, \bibinfo {author} {\bibfnamefont {G.~L.}\
  \bibnamefont {Chiarotti}}, \bibinfo {author} {\bibfnamefont {M.}~\bibnamefont
  {Cococcioni}}, \bibinfo {author} {\bibfnamefont {I.}~\bibnamefont {Dabo}},
  \emph {et~al.},\ }\href@noop {} {\bibfield  {journal} {\bibinfo  {journal}
  {Journal of physics: Condensed matter}\ }\textbf {\bibinfo {volume} {21}},\
  \bibinfo {pages} {395502} (\bibinfo {year} {2009})}\BibitemShut {NoStop}%
\bibitem [{\citenamefont {Coelho}\ \emph {et~al.}(2019)\citenamefont {Coelho},
  \citenamefont {Nguyen~Cong}, \citenamefont {Bonilla}, \citenamefont
  {Kolekar}, \citenamefont {Phan}, \citenamefont {Avila}, \citenamefont
  {Asensio}, \citenamefont {Oleynik},\ and\ \citenamefont
  {Batzill}}]{coelho2019charge}%
  \BibitemOpen
  \bibfield  {author} {\bibinfo {author} {\bibfnamefont {P.~M.}\ \bibnamefont
  {Coelho}}, \bibinfo {author} {\bibfnamefont {K.}~\bibnamefont {Nguyen~Cong}},
  \bibinfo {author} {\bibfnamefont {M.}~\bibnamefont {Bonilla}}, \bibinfo
  {author} {\bibfnamefont {S.}~\bibnamefont {Kolekar}}, \bibinfo {author}
  {\bibfnamefont {M.-H.}\ \bibnamefont {Phan}}, \bibinfo {author}
  {\bibfnamefont {J.}~\bibnamefont {Avila}}, \bibinfo {author} {\bibfnamefont
  {M.~C.}\ \bibnamefont {Asensio}}, \bibinfo {author} {\bibfnamefont {I.~I.}\
  \bibnamefont {Oleynik}}, \ and\ \bibinfo {author} {\bibfnamefont
  {M.}~\bibnamefont {Batzill}},\ }\href@noop {} {\bibfield  {journal} {\bibinfo
   {journal} {The Journal of Physical Chemistry C}\ }\textbf {\bibinfo {volume}
  {123}},\ \bibinfo {pages} {14089} (\bibinfo {year} {2019})}\BibitemShut
  {NoStop}%
\bibitem [{\citenamefont {Feng}\ \emph {et~al.}(2018)\citenamefont {Feng},
  \citenamefont {Biswas}, \citenamefont {Rajan}, \citenamefont {Watson},
  \citenamefont {Mazzola}, \citenamefont {Clark}, \citenamefont {Underwood},
  \citenamefont {Markovic}, \citenamefont {McLaren}, \citenamefont {Hunter}
  \emph {et~al.}}]{feng2018electronic}%
  \BibitemOpen
  \bibfield  {author} {\bibinfo {author} {\bibfnamefont {J.}~\bibnamefont
  {Feng}}, \bibinfo {author} {\bibfnamefont {D.}~\bibnamefont {Biswas}},
  \bibinfo {author} {\bibfnamefont {A.}~\bibnamefont {Rajan}}, \bibinfo
  {author} {\bibfnamefont {M.~D.}\ \bibnamefont {Watson}}, \bibinfo {author}
  {\bibfnamefont {F.}~\bibnamefont {Mazzola}}, \bibinfo {author} {\bibfnamefont
  {O.~J.}\ \bibnamefont {Clark}}, \bibinfo {author} {\bibfnamefont
  {K.}~\bibnamefont {Underwood}}, \bibinfo {author} {\bibfnamefont
  {I.}~\bibnamefont {Markovic}}, \bibinfo {author} {\bibfnamefont
  {M.}~\bibnamefont {McLaren}}, \bibinfo {author} {\bibfnamefont
  {A.}~\bibnamefont {Hunter}},  \emph {et~al.},\ }\href@noop {} {\bibfield
  {journal} {\bibinfo  {journal} {Nano letters}\ }\textbf {\bibinfo {volume}
  {18}},\ \bibinfo {pages} {4493} (\bibinfo {year} {2018})}\BibitemShut
  {NoStop}%
\bibitem [{\citenamefont {Chen}\ \emph {et~al.}(2018)\citenamefont {Chen},
  \citenamefont {Pai}, \citenamefont {Chan}, \citenamefont {Madhavan},
  \citenamefont {Chou}, \citenamefont {Mo}, \citenamefont {Fedorov},\ and\
  \citenamefont {Chiang}}]{chen2018unique}%
  \BibitemOpen
  \bibfield  {author} {\bibinfo {author} {\bibfnamefont {P.}~\bibnamefont
  {Chen}}, \bibinfo {author} {\bibfnamefont {W.~W.}\ \bibnamefont {Pai}},
  \bibinfo {author} {\bibfnamefont {Y.-H.}\ \bibnamefont {Chan}}, \bibinfo
  {author} {\bibfnamefont {V.}~\bibnamefont {Madhavan}}, \bibinfo {author}
  {\bibfnamefont {M.-Y.}\ \bibnamefont {Chou}}, \bibinfo {author}
  {\bibfnamefont {S.-K.}\ \bibnamefont {Mo}}, \bibinfo {author} {\bibfnamefont
  {A.-V.}\ \bibnamefont {Fedorov}}, \ and\ \bibinfo {author} {\bibfnamefont
  {T.-C.}\ \bibnamefont {Chiang}},\ }\href@noop {} {\bibfield  {journal}
  {\bibinfo  {journal} {Physical review letters}\ }\textbf {\bibinfo {volume}
  {121}},\ \bibinfo {pages} {196402} (\bibinfo {year} {2018})}\BibitemShut
  {NoStop}%
\bibitem [{\citenamefont {Kang}\ \emph {et~al.}(2003)\citenamefont {Kang},
  \citenamefont {Kim}, \citenamefont {Sekiyama}, \citenamefont {Kasai},
  \citenamefont {Suga}, \citenamefont {Han}, \citenamefont {Kim}, \citenamefont
  {Choi}, \citenamefont {Kimura}, \citenamefont {Muro}, \citenamefont {Saitoh},
  \citenamefont {Olson}, \citenamefont {Shim},\ and\ \citenamefont
  {Min}}]{kang2003resonant}%
  \BibitemOpen
  \bibfield  {author} {\bibinfo {author} {\bibfnamefont {J.~S.}\ \bibnamefont
  {Kang}}, \bibinfo {author} {\bibfnamefont {J.~H.}\ \bibnamefont {Kim}},
  \bibinfo {author} {\bibfnamefont {A.}~\bibnamefont {Sekiyama}}, \bibinfo
  {author} {\bibfnamefont {S.}~\bibnamefont {Kasai}}, \bibinfo {author}
  {\bibfnamefont {S.}~\bibnamefont {Suga}}, \bibinfo {author} {\bibfnamefont
  {S.~W.}\ \bibnamefont {Han}}, \bibinfo {author} {\bibfnamefont {K.~H.}\
  \bibnamefont {Kim}}, \bibinfo {author} {\bibfnamefont {E.~J.}\ \bibnamefont
  {Choi}}, \bibinfo {author} {\bibfnamefont {T.}~\bibnamefont {Kimura}},
  \bibinfo {author} {\bibfnamefont {T.}~\bibnamefont {Muro}}, \bibinfo {author}
  {\bibfnamefont {Y.}~\bibnamefont {Saitoh}}, \bibinfo {author} {\bibfnamefont
  {C.~G.}\ \bibnamefont {Olson}}, \bibinfo {author} {\bibfnamefont {J.~H.}\
  \bibnamefont {Shim}}, \ and\ \bibinfo {author} {\bibfnamefont {B.~I.}\
  \bibnamefont {Min}},\ }\href@noop {} {\bibfield  {journal} {\bibinfo
  {journal} {Physical Review B}\ }\textbf {\bibinfo {volume} {68}},\ \bibinfo
  {pages} {012410} (\bibinfo {year} {2003})}\BibitemShut {NoStop}%
\bibitem [{\citenamefont {Riley}\ \emph {et~al.}(2015)\citenamefont {Riley},
  \citenamefont {Meevasana}, \citenamefont {Bawden}, \citenamefont {Asakawa},
  \citenamefont {Takayama}, \citenamefont {Eknapakul}, \citenamefont {Kim},
  \citenamefont {Hoesch}, \citenamefont {Mo}, \citenamefont {Takagi} \emph
  {et~al.}}]{riley2015negative}%
  \BibitemOpen
  \bibfield  {author} {\bibinfo {author} {\bibfnamefont {J.~M.}\ \bibnamefont
  {Riley}}, \bibinfo {author} {\bibfnamefont {W.}~\bibnamefont {Meevasana}},
  \bibinfo {author} {\bibfnamefont {L.}~\bibnamefont {Bawden}}, \bibinfo
  {author} {\bibfnamefont {M.}~\bibnamefont {Asakawa}}, \bibinfo {author}
  {\bibfnamefont {T.}~\bibnamefont {Takayama}}, \bibinfo {author}
  {\bibfnamefont {T.}~\bibnamefont {Eknapakul}}, \bibinfo {author}
  {\bibfnamefont {T.~K.}\ \bibnamefont {Kim}}, \bibinfo {author} {\bibfnamefont
  {M.}~\bibnamefont {Hoesch}}, \bibinfo {author} {\bibfnamefont {S.-K.}\
  \bibnamefont {Mo}}, \bibinfo {author} {\bibfnamefont {H.}~\bibnamefont
  {Takagi}},  \emph {et~al.},\ }\href@noop {} {\bibfield  {journal} {\bibinfo
  {journal} {Nature nanotechnology}\ }\textbf {\bibinfo {volume} {10}},\
  \bibinfo {pages} {1043} (\bibinfo {year} {2015})}\BibitemShut {NoStop}%
\bibitem [{\citenamefont {Kang}\ \emph {et~al.}(2017)\citenamefont {Kang},
  \citenamefont {Kim}, \citenamefont {Ryu}, \citenamefont {Jung}, \citenamefont
  {Kim}, \citenamefont {Moreschini}, \citenamefont {Jozwiak}, \citenamefont
  {Rotenberg}, \citenamefont {Bostwick},\ and\ \citenamefont
  {Kim}}]{kang2017universal}%
  \BibitemOpen
  \bibfield  {author} {\bibinfo {author} {\bibfnamefont {M.}~\bibnamefont
  {Kang}}, \bibinfo {author} {\bibfnamefont {B.}~\bibnamefont {Kim}}, \bibinfo
  {author} {\bibfnamefont {S.~H.}\ \bibnamefont {Ryu}}, \bibinfo {author}
  {\bibfnamefont {S.~W.}\ \bibnamefont {Jung}}, \bibinfo {author}
  {\bibfnamefont {J.}~\bibnamefont {Kim}}, \bibinfo {author} {\bibfnamefont
  {L.}~\bibnamefont {Moreschini}}, \bibinfo {author} {\bibfnamefont
  {C.}~\bibnamefont {Jozwiak}}, \bibinfo {author} {\bibfnamefont
  {E.}~\bibnamefont {Rotenberg}}, \bibinfo {author} {\bibfnamefont
  {A.}~\bibnamefont {Bostwick}}, \ and\ \bibinfo {author} {\bibfnamefont
  {K.~S.}\ \bibnamefont {Kim}},\ }\href@noop {} {\bibfield  {journal} {\bibinfo
   {journal} {Nano letters}\ }\textbf {\bibinfo {volume} {17}},\ \bibinfo
  {pages} {1610} (\bibinfo {year} {2017})}\BibitemShut {NoStop}%
\bibitem [{\citenamefont {Li}\ \emph {et~al.}(2021)\citenamefont {Li},
  \citenamefont {Ren},\ and\ \citenamefont {Shuai}}]{li2021general}%
  \BibitemOpen
  \bibfield  {author} {\bibinfo {author} {\bibfnamefont {W.}~\bibnamefont
  {Li}}, \bibinfo {author} {\bibfnamefont {J.}~\bibnamefont {Ren}}, \ and\
  \bibinfo {author} {\bibfnamefont {Z.}~\bibnamefont {Shuai}},\ }\href@noop {}
  {\bibfield  {journal} {\bibinfo  {journal} {Nature Communications}\ }\textbf
  {\bibinfo {volume} {12}},\ \bibinfo {pages} {1} (\bibinfo {year}
  {2021})}\BibitemShut {NoStop}%
\bibitem [{\citenamefont {Fetherolf}\ \emph {et~al.}(2020)\citenamefont
  {Fetherolf}, \citenamefont {Gole{\v{z}}},\ and\ \citenamefont
  {Berkelbach}}]{fetherolf2020unification}%
  \BibitemOpen
  \bibfield  {author} {\bibinfo {author} {\bibfnamefont {J.~H.}\ \bibnamefont
  {Fetherolf}}, \bibinfo {author} {\bibfnamefont {D.}~\bibnamefont
  {Gole{\v{z}}}}, \ and\ \bibinfo {author} {\bibfnamefont {T.~C.}\ \bibnamefont
  {Berkelbach}},\ }\href@noop {} {\bibfield  {journal} {\bibinfo  {journal}
  {Physical Review X}\ }\textbf {\bibinfo {volume} {10}},\ \bibinfo {pages}
  {021062} (\bibinfo {year} {2020})}\BibitemShut {NoStop}%
\bibitem [{\citenamefont {Berciu}\ and\ \citenamefont
  {Sawatzky}(2008)}]{berciu2008light}%
  \BibitemOpen
  \bibfield  {author} {\bibinfo {author} {\bibfnamefont {M.}~\bibnamefont
  {Berciu}}\ and\ \bibinfo {author} {\bibfnamefont {G.~A.}\ \bibnamefont
  {Sawatzky}},\ }\href@noop {} {\bibfield  {journal} {\bibinfo  {journal} {EPL
  (Europhysics Letters)}\ }\textbf {\bibinfo {volume} {81}},\ \bibinfo {pages}
  {57008} (\bibinfo {year} {2008})}\BibitemShut {NoStop}%
\bibitem [{\citenamefont {Kang}\ \emph {et~al.}(2018)\citenamefont {Kang},
  \citenamefont {Jung}, \citenamefont {Shin}, \citenamefont {Sohn},
  \citenamefont {Ryu}, \citenamefont {Kim}, \citenamefont {Hoesch},\ and\
  \citenamefont {Kim}}]{kang2018holstein}%
  \BibitemOpen
  \bibfield  {author} {\bibinfo {author} {\bibfnamefont {M.}~\bibnamefont
  {Kang}}, \bibinfo {author} {\bibfnamefont {S.~W.}\ \bibnamefont {Jung}},
  \bibinfo {author} {\bibfnamefont {W.~J.}\ \bibnamefont {Shin}}, \bibinfo
  {author} {\bibfnamefont {Y.}~\bibnamefont {Sohn}}, \bibinfo {author}
  {\bibfnamefont {S.~H.}\ \bibnamefont {Ryu}}, \bibinfo {author} {\bibfnamefont
  {T.~K.}\ \bibnamefont {Kim}}, \bibinfo {author} {\bibfnamefont
  {M.}~\bibnamefont {Hoesch}}, \ and\ \bibinfo {author} {\bibfnamefont {K.~S.}\
  \bibnamefont {Kim}},\ }\href@noop {} {\bibfield  {journal} {\bibinfo
  {journal} {Nature materials}\ }\textbf {\bibinfo {volume} {17}},\ \bibinfo
  {pages} {676} (\bibinfo {year} {2018})}\BibitemShut {NoStop}%
\bibitem [{\citenamefont {Sahoo}\ \emph {et~al.}(2020)\citenamefont {Sahoo},
  \citenamefont {Dutta}, \citenamefont {Harnagea}, \citenamefont {Sood},\ and\
  \citenamefont {Karmakar}}]{sahoo2020pressure}%
  \BibitemOpen
  \bibfield  {author} {\bibinfo {author} {\bibfnamefont {S.}~\bibnamefont
  {Sahoo}}, \bibinfo {author} {\bibfnamefont {U.}~\bibnamefont {Dutta}},
  \bibinfo {author} {\bibfnamefont {L.}~\bibnamefont {Harnagea}}, \bibinfo
  {author} {\bibfnamefont {A.~K.}\ \bibnamefont {Sood}}, \ and\ \bibinfo
  {author} {\bibfnamefont {S.}~\bibnamefont {Karmakar}},\ }\href@noop {}
  {\bibfield  {journal} {\bibinfo  {journal} {Physical Review B}\ }\textbf
  {\bibinfo {volume} {101}},\ \bibinfo {pages} {014514} (\bibinfo {year}
  {2020})}\BibitemShut {NoStop}%
\bibitem [{\citenamefont {Mannella}\ \emph {et~al.}(2007)\citenamefont
  {Mannella}, \citenamefont {Yang}, \citenamefont {Tanaka}, \citenamefont
  {Zhou}, \citenamefont {Zheng}, \citenamefont {Mitchell}, \citenamefont
  {Zaanen}, \citenamefont {Devereaux}, \citenamefont {Nagaosa}, \citenamefont
  {Hussain},\ and\ \citenamefont {Shen}}]{mannella2007polaron}%
  \BibitemOpen
  \bibfield  {author} {\bibinfo {author} {\bibfnamefont {N.}~\bibnamefont
  {Mannella}}, \bibinfo {author} {\bibfnamefont {W.~L.}\ \bibnamefont {Yang}},
  \bibinfo {author} {\bibfnamefont {K.}~\bibnamefont {Tanaka}}, \bibinfo
  {author} {\bibfnamefont {X.~J.}\ \bibnamefont {Zhou}}, \bibinfo {author}
  {\bibfnamefont {H.}~\bibnamefont {Zheng}}, \bibinfo {author} {\bibfnamefont
  {J.~F.}\ \bibnamefont {Mitchell}}, \bibinfo {author} {\bibfnamefont
  {J.}~\bibnamefont {Zaanen}}, \bibinfo {author} {\bibfnamefont {T.~P.}\
  \bibnamefont {Devereaux}}, \bibinfo {author} {\bibfnamefont {N.}~\bibnamefont
  {Nagaosa}}, \bibinfo {author} {\bibfnamefont {Z.}~\bibnamefont {Hussain}}, \
  and\ \bibinfo {author} {\bibfnamefont {Z.-X.}\ \bibnamefont {Shen}},\
  }\href@noop {} {\bibfield  {journal} {\bibinfo  {journal} {Physical Review
  B}\ }\textbf {\bibinfo {volume} {76}},\ \bibinfo {pages} {233102} (\bibinfo
  {year} {2007})}\BibitemShut {NoStop}%
\end{thebibliography}%

\end{document}